\documentclass[doublecol]{epl2}
\usepackage{epstopdf}
\title{Magnetic relaxation and collective vortex creep in FeTe$_{0.6}$Se$_{0.4}$ single crystal}
\shorttitle{Title} %Insert here a short version of the title if it exceeds 70 characters

\author{Yue Sun\inst{1,2} \and Toshihiro Taen\inst{2} \and Yuji Tsuchiya\inst{2} \and Sunseng Pyon\inst{2} \and Zhixiang Shi\inst{1*} \and Tsuyoshi Tamegai\inst{2\dagger}}
\shortauthor{Yue Sun \etal}

\institute{
  \inst{1} Department of Physics, Southeast University, Nanjing 211189, People¡¯s Republic of China\\
  \inst{2} Department of Applied Physics, The University of Tokyo, 7-3-1 Hongo, Bunkyo-ku, Tokyo 113-8656, Japan
}
\pacs{74.25.Wx}{Vortex pinning}
\pacs{74.25.Sv}{Critical currents}
\pacs{74.70.Xa}{Pnictides and chalcogenides}

\abstract{
We study the vortex dynamics in high-quality FeTe$_{0.6}$Se$_{0.4}$ single crystal by performing magnetization measurements of the screening current density \emph{J}$_s$ and flux creep rate \emph{S}. Temperature dependence of \emph{S} shows a plateau in the intermediate temperature region with a high creep rate $\sim$ 0.03, which is interpreted in the framework of the collective creep theory. A crossover from elastic to plastic creep is observed. The glassy exponent and barrier height for flux creep are directly determined by extended Maley's method. \emph{J}$_s$ with flux creep, obtained from magnetic hysteresis loops, is successfully reproduced based on the collective creep analysis. We also approach critical current density without flux creep by means of the generalized inversion scheme, which proves that the $\delta$\emph{l} and $\delta$\emph{T}$_c$ pinning coexist in FeTe$_{0.6}$Se$_{0.4}$ single crystal. }

\begin{document}

\maketitle

Recently discovered iron-based superconductors (IBSs) with superconducting transition temperature \emph{T}$_c$ above 55 K is another member of the high temperature superconductors (HTS) after cuprate superconductors \cite{1,2}. IBSs share some similarities with cuprate superconductors like layered structure, very high upper critical fields and unconventional paring mechanism \cite{2,3}. The study of IBSs is helpful to solve some puzzles remained in HTS and testify some notions and theories originated from the cuprate superconductors. Among these, vortex dynamics is one of the central issues related to both basic science and technological applications.  For cuprate superconductors, due to large anisotropy, short coherence lengths and high operation temperature, the vortex motion and fluctuations are quite strong \cite{4}. This leads to a collective pinning with small characteristic pinning energy, which give rise to intriguing experimental results, such as ¡°plateau¡± observed in the temperature dependence of normalized relaxation rate (\emph{S}$\equiv$$\mid$dln\emph{M}/dln\emph{t}$\mid$) \cite{5} in contrast to the linear increase with temperature predicted by the Anderson-Kim model in low temperature superconductors (LTS). To understand these behaviors, some theories have been proposed in the past decades \cite{4}. Among these, the collective creep theory successfully interpreted the ¡°plateau¡± region, and the large creep rate \cite{5,6}. The discovery of IBSs provides another opportunity to study the vortex dynamics as well as collective creep theory in HTS, and its intermediate \emph{T}$_c$ is also meaningful to understand the crossover between LTS and HTS.

Magnetic relaxation measurements in IBSs reported so far have revealed that they also show giant flux creep and collective pinning, which implies that IBSs and cuprate superconductors may share common vortex physics \cite{7,8,9,10,11,12,13}. Until now, most detailed studies about the vortex dynamics on IBSs have been performed on "122" phase since high-quality single crystals are readily available. Especially in Co-doped BaFe$_2$As$_2$, it shows fast flux creep and a transition from collective to plastic creep \cite{7,9,11,12}. In order to know if the large creep rate governed by collective flux creep is an intrinsic property of IBSs, detailed vortex dynamics studies should be done on other IBS samples. Among IBSs, FeTe$_{1-\emph{x}}$Se$_{x}$ composed of only Fe(Te,Se) layers is a preferred choice to study vortex dynamics because of its simple crystal structure. On the other hand, its less toxic nature makes FeTe$_{1-\emph{x}}$Se$_{x}$ a more suitable candidate for applications among the family of IBSs. However, although large single crystals can be easily obtained, the existence of excess Fe affects the sample quality leading to an inhomogeneous distribution of \emph{T}$_c$ and \emph{J}$_c$ \cite{14,15}. Thus only a limited number of works have been done on the vortex dynamics in iron chalcogenides, and some reported results are still in controversy because of the difference in sample quality \cite{16,17,18,19,20}. Furthermore, there is no direct evidence of the collective pinning like the ¡°plateau¡± in the temperature dependence of relaxation rate has been reported.

\begin{figure}\center
\onefigure[width=8.5cm]{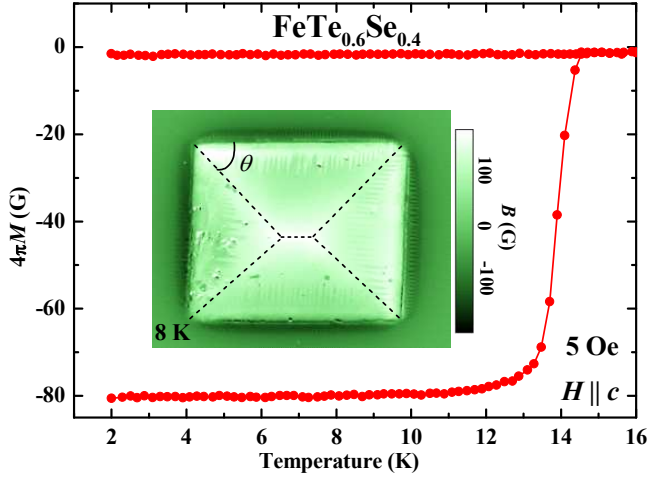}
\caption{Temperature dependences of zero-field-cooled (ZFC) and field-cooled (FC) magnetizations at 5 Oe for FeTe$_{0.6}$Se$_{0.4}$ single crystal. The inset shows magneto-optical image in the remanent state after applying 400 Oe along the \emph{c}-axis at 8 K.}
\label{fig.1}
\end{figure}

Recently, we developed a controllable way of removing the excess Fe by annealing with a controlled amount of O$_2$, and obtained a high-quality single crystal with large and homogeneous \emph{J}$_c$ \cite{21,22}. In this paper, we report a detailed study of vortex dynamics in a well annealed FeTe$_{0.6}$Se$_{0.4}$ single crystal. Temperature dependence of \emph{S} shows a plateau in the intermediate temperature region with a large vortex creep rate \emph{S} $\sim$ 0.03, which can be interpreted in the framework of collective creep theory. A crossover from elastic to plastic creep was observed. Screening current density with flux creep, obtained from magnetic hysteresis loops (MHLs), is successfully reproduced based on the collective creep analysis. We also approach critical current density without flux creep by means of the generalized inversion scheme, which proves that the $\delta$\emph{l} and $\delta$\emph{T}$_c$ pinning coexist in FeTe$_{0.6}$Se$_{0.4}$ single crystals.

Single crystals with a nominal composition FeTe$_{0.6}$Se$_{0.4}$ were grown by the self-flux method. The obtained as-grown crystals were further annealed with a fixed amount of O$_2$ to induce bulk and homogeneous superconductivity. Details of the crystal growth and O$_2$-annealing have been reported in our previous publications \cite{21,22,23}. Magnetization measurements were performed using a commercial SQUID magnetometer (MPMS-XL5, Quantum Design). Magneto-optical (MO) images were obtained by using the local field-dependent Faraday effect in the in-plane magnetized garnet indicator film employing a differential method \cite{24,25}. Screening current density calculated by the Bean model is denoted as \emph{J}$_s$ in field-sweep measurements or simply \emph{J} in relaxation measurements.

Temperature dependence of magnetization was measured to check the quality of FeTe$_{0.6}$Se$_{0.4}$ single crystal. As shown in the main panel of Fig. 1, the crystal displays a superconducting transition temperature, \emph{T}$_c$ $\sim$ 14.5 K with a transition width less than 1 K (obtained from the criteria of 10 and 90\% of the magnetization result at 2 K). To further confirm the homogeneity of the crystal, we took MO images on the same piece of single crystal in the remanent state. A typical MO image taken at 8 K after cycling the field up to 400 Oe along \emph{c}-axis is shown in the inset of Fig. 1. The MO image manifests a typical roof-top pattern, indicating a nearly uniform current flow in the crystal. Besides, the typical current discontinuity lines (so-called \emph{d}-line), which cannot be crossed by vortices, can be observed and marked by the dotted line. The angle $\theta$ of the \emph{d}-line for our rectangular sample is $\sim$ 45$^\circ$, indicating that the critical current density within the \emph{ab} plane is isotropic, consistent with the four-fold symmetry of the superconducting plane. The homogeneous current flow within the sample proved by the MO result also guarantees the reliability of using the Bean model to estimate \emph{J}$_s$ from MHLs.

\begin{figure}\center
\onefigure[width=8.5cm]{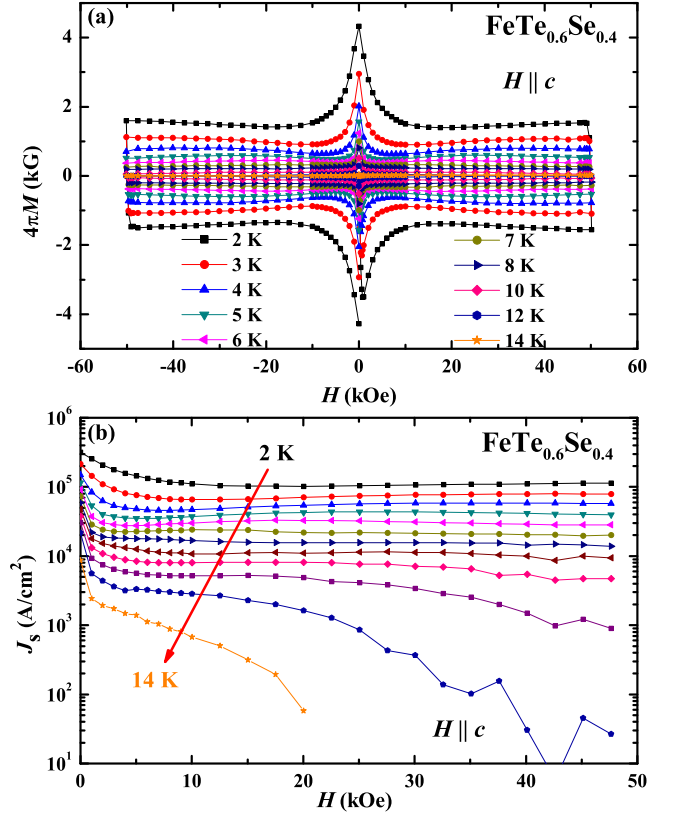}
\caption{(a) Magnetic hysteresis loops of FeTe$_{0.6}$Se$_{0.4}$ at different temperatures ranging from 2 to 14 K for \emph{H} $\|$ \emph{c}. (b) Magnetic field dependence of critical current densities for \emph{H} $\|$ \emph{c}.}
\label{fig.2}
\end{figure}

Fig. 2(a) shows the MHLs obtained from the same piece of crystal at temperatures ranging from 2 to 14 K for \emph{H} $\|$ \emph{c}. A second magnetization peak (SMP), also known as the fish-tail effect (FE), can be witnessed, which is a common feature of iron-based superconductors. The SMP can be witnessed more clearly in the field dependent critical current density in Fig. 2(b) obtained by using the Bean model \cite{26}:
\begin{equation}
\label{eq.1}
J_c=20\frac{\Delta M}{a(1-a/3b)},
\end{equation}
where $\Delta$\emph{M} is \emph{M}$_{down}$ - \emph{M}$_{up}$, \emph{M}$_{up}$ [emu/cm$^3$] and \emph{M}$_{down}$ [emu/cm$^3$] are the magnetization when sweeping fields up and down, respectively, \emph{a} [cm] and \emph{b} [cm] are sample widths (\emph{a} $<$ \emph{b}). Here we should point out that the \emph{J}$_s$ calculated from the MHLs is the critical current density with flux creep, since there is a finite time delay between the measurement and the preparation of critical state. The self-field \emph{J}$_s$ reaches a large value about 3 $\times$ 10$^5$ A/cm$^2$ at 2 K, and keeps a value $\sim$ 1 $\times$ 10$^5$ A/cm$^2$ even at 50 kOe. Although the value of \emph{J}$_s$ is lower than that of Ba(Fe$_{1-\emph{x}}$Co$_\emph{x}$)$_2$As$_2$ single crystal \cite{27}, it is one of the largest values among those reported in Fe(Te,Se) \cite{21}. The large value of \emph{J}$_s$ again manifests the high quality of the crystal and ensures that the vortex dynamics study probes the intrinsic property of FeTe$_{0.6}$Se$_{0.4}$ single crystal.

\begin{figure}\center
\onefigure[width=8.5cm]{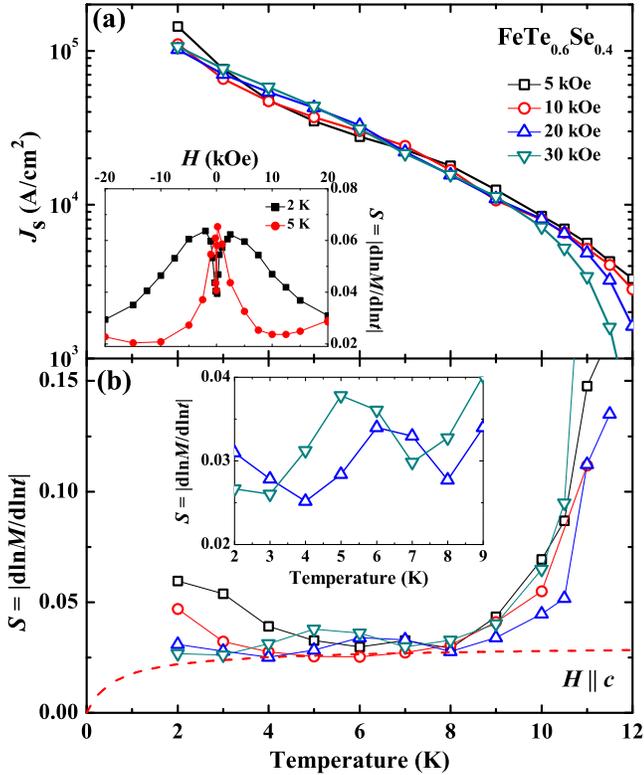}
\caption{(a) Temperature dependences of log\emph{J}$_s$ at applied fields ranging from 5 to 30 kOe. The inset shows field dependence of the magnetic relaxation rate at 2 and 5 K. (b) Temperature dependence of \emph{S} at applied fields ranging from 5 to 30 kOe. The inset is the enlarge part of \emph{S} at 20 and 30 kOe from 2 to 9 K. }
\label{fig.3}
\end{figure}

Fig. 3(b) shows the temperature dependence of magnetic relaxation rate \emph{S}$\equiv$$\mid$dln\emph{M}/dln\emph{t}$\mid$ at applied field ranging from 5 to 30 kOe, where \emph{M} is magnetization and \emph{t} is time from the moment when the critical state is prepared. In these measurements, magnetic field was swept more than 5 kOe higher before decreasing to the target field. Obviously, there is a plateau in the intermediate temperature range with a high vortex creep rate \emph{S} $\sim$ 0.03. The plateau and large vortex creep rate were also observed in YBa$_2$Cu$_3$O$_{7-\emph{$\delta$}}$ \cite{6}, and iron-based "122" \cite{7,11,12,28} and "1111" samples \cite{13}, which can be interpreted by the collective creep theory \cite{6}. On the other hand, \emph{S} - \emph{T} curve undergoes an upturn at low temperatures especially at small fields like 5 and 10 kOe. In the context of collective creep theory, the increase of \emph{S} corresponds to the smaller value of \emph{$\mu$}, suggesting that, at least, a part of the vortex system approaches to the single vortex regime with \emph{$\mu$} = 1/7. As the temperature is lowered, \emph{J}$_c$ increases leading to the wider distribution of the local field in the sample. When the applied field is not considerably larger than the self-field, local magnetic induction in the region close to the edge of the sample becomes much smaller than the applied field, making this region close to the single vortex regime with smaller \emph{$\mu$}. On the other hand, when the applied field becomes much larger than the self-field, no regions in the sample can experience low enough field for the single vortex regime. Hence, the increase of \emph{S} becomes not obvious at higher fields like 20 or 30 kOe. Actually it can be seen more clearly in the field dependence of magnetic relaxation rate (\emph{S} - \emph{H}) as shown in the inset of Fig. 3(a). \emph{S} shows an increase with decreasing field, while it drops down suddenly below the self-field forming a dip in \emph{S} - \emph{H}. According to Taen \emph{et al}. \cite{17}, this behavior may occur when the applied field \emph{H} is smaller than the self-field being a consequence of the ¡°Meissner hole¡± that appears close to the edges of the sample \cite{29}. Similar dip structure of field dependent \emph{S} have been reported in Bi$_2$Sr$_2$CaCu$_2$O$_{8+\emph{y}}$ [30] and Ba(Fe$_{1-\emph{x}}$Co$_\emph{x}$)$_2$As$_2$ \cite{11,27}.

Another notable feature in the temperature dependence of \emph{S} in FeTe$_{0.6}$Se$_{0.4}$ is the hump in the intermediate temperature range for fields between 20 and 30 kOe, which can be seen more clearly in the inset of Fig. 3(b). This hump-like behavior is related to the SMP observed in the MHLs. To compare the hump-like behavior with the SMP, we also plotted the log\emph{J}$_s$ - \emph{T} curves in the same magnetic field and temperature range in the main panel of Fig. 3(a). It is clear that \emph{J}$_s$'s under different fields show crossovers between $\sim$ 3 K and $\sim$ 7 K. In this temperature range, \emph{J}$_s$ at high field is larger than that at low fields, which comes from the SMP. Actually, this temperature range is exactly the same range where \emph{S} manifests a hump-like behavior. A similar effect of SMP on the magnetic relaxation rate was also reported in Co-doped BaFe$_2$As$_2$ \cite{7,11} and Na-doped CaFe$_2$As$_2$ \cite{28}.

\begin{figure}\center
\onefigure[width=8.5cm]{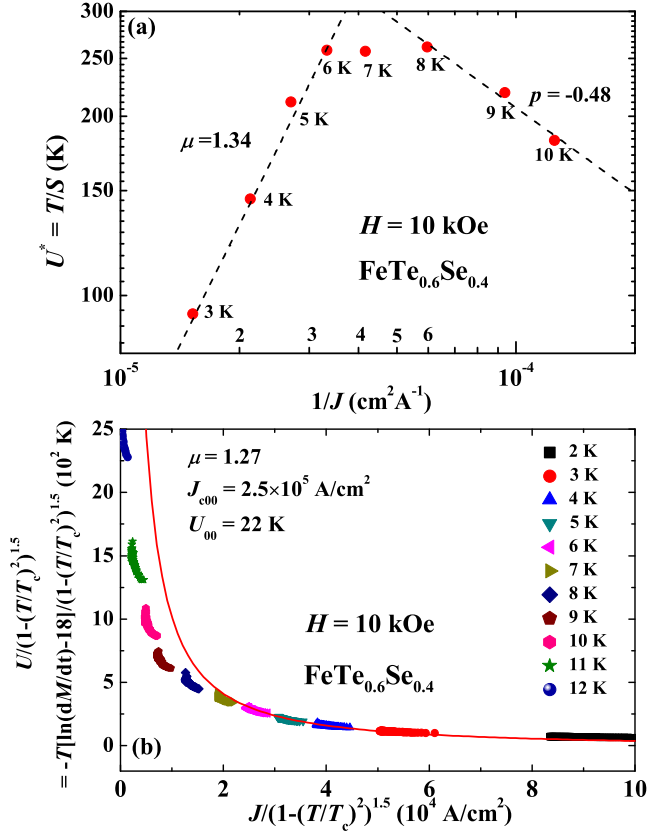}
\caption{(a) Inverse current density dependence of effective pinning energy \emph{U}$^*$ at 10 kOe in FeTe$_{0.6}$Se$_{0.4}$ single crystal. (b) Current density dependence of flux activation energy \emph{U} constructed by the extended Maley's method. The solid line indicates power-law fitting in the large \emph{J} region. }
\label{fig.4}
\end{figure}

Based on the discussion above, we chose the \emph{S} - \emph{T} data at 10 kOe, which is little affected by SMP and by the self-field effect compared to the case of 5 kOe to further probe the vortex dynamics of FeTe$_{0.6}$Se$_{0.4}$ single crystal. According to the collective creep theory \cite{6}, magnetic relaxation rate \emph{S} can be described as
\begin{equation}
\label{eq.2}
S=\frac{T}{U_0+\mu Tln(t/t_{eff})},
\end{equation}
where \emph{U}$_0$ is the temperature-dependent flux activation energy in the absence of flux creep, \emph{t}$_{eff}$ is the effective hopping attempt time, and $\mu$ $>$ 0 is a glassy exponent for elastic creep. The value of $\mu$ contains information about the size of the vortex bundle in the collective creep theory. In a three-dimensional system, it is predicted as $\mu$ = 1/7, (1) 5/2, 7/9 for single-vortex, (intermadiate) small-bundle, and large-bundle regimes, respectively \cite{4,31}.

The flux activation energy \emph{U} as a function of current density \emph{J} can be defined as \cite{32}
\begin{equation}
\label{eq.3}
U(J)=\frac{U_0}{\mu}[(J_{c0}/J)^\mu-1].
\end{equation}
Combining this with \emph{U} = \emph{T}ln(\emph{t}/\emph{t}$_{eff}$) extracted from the Arrhenius relation, we can deduce the so-called ¡°interpolation formula¡±
\begin{equation}
\label{eq.4}
J(T,t)=\frac{J_{c0}}{[1+(\mu T/U_0)ln(t/t_{eff})]^{1/\mu}},
\end{equation}
where \emph{J}$_{c0}$ is the temperature dependent critical current density in the absence of flux creep. From Eqs. (3) and (4), effective pinning energy \emph{U}$^*$ = \emph{T} / \emph{S} can be derived as
\begin{equation}
\label{eq.5}
U^*=U_0+\mu Tln(t/t_{eff})=U_0(J_{c0}/J)^\mu.
\end{equation}
\begin{figure}\center
\onefigure[width=8.5cm]{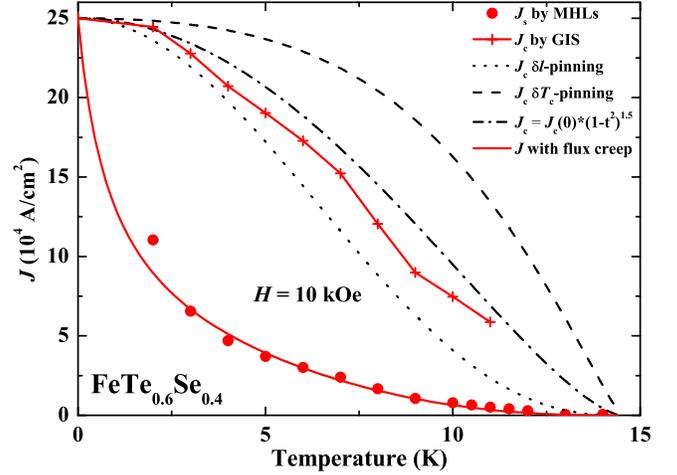}
\caption{Temperature dependence of \emph{J}$_s$ determined from the MHLs, model function of \emph{J} before and after creep by the collective creep theory, \emph{J}$_c$ calculated from $\delta$\emph{l} and $\delta$\emph{T}$_c$ pinning, and \emph{J}$_c$ without flux creep reconstructed by GIS. }
\label{fig.5}
\end{figure}
Thus, the value of $\mu$ can be easily obtained from the slope in the double logarithmic plot of \emph{U}$^*$ vs 1/\emph{J}, as shown in the Fig. 4(a). The evaluated value of $\mu$ is $\sim$ 1.34 close to that reported in YBa$_2$Cu$_3$O$_{7-\emph{$\delta$}}$ \cite{33}, and IBSs \cite{10,12,20}. The value of $\mu$ resides between the prediction of single-vortex (1/7) and small-bundle (5/2) regimes, indicating that contributions of those two kinds of pinnings coexist. Usually, at low field and low temperature region, the flux creep is dominated by the motion of individual flux lines without interaction. When temperature and field are increased, the interaction between flux lines becomes non-negligible, and they will creep collectively in the form of small (or intermediate) bundles. However, it is very difficult to determine the boundary of these two regimes \cite{4}. In our case, \emph{S} increases at low temperature and small field as shown in Fig. 3, which means the single-vortex motion becomes more dominant. Contrary to the above prediction of $\mu$ $>$ 0, a negative slope \emph{p} = -0.48 is obtained at small \emph{J}, which is very close to the value of -0.5 predicted by plastic creep theory \cite{34}. Thus the temperature dependence of \emph{S} shows a crossover between elastic and plastic creep regimes.

In the following, we analyze the \emph{U}(\emph{J}) relation by the extended Maley's method \cite{35}, which considering the temperature dependence of \emph{U} and \emph{J} into the original Maley's method \cite{35}. This method allows to scale \emph{U} in a wide range of \emph{J}. The temperature dependent \emph{U}$_0$ and \emph{J}$_{c0}$ is assumed as
\begin{equation}
\label{eq.6}
U_0(T)=U_{00}[1-(T/T_c)^2]^n.
\end{equation}
\begin{equation}
\label{eq.7}
J_{c0}(T)=J_{c00}[1-(T/T_c)^2]^n.
\end{equation}
Here, the exponent \emph{n} is set to 3/2 as in the case of  YBa$_2$Cu$_3$O$_{7-\emph{$\delta$}}$ \cite{33,35} and Co-doped BaFe$_2$As$_2$ \cite{12}.\emph{U} = -\emph{T}ln[d\emph{M}(\emph{t})/d(\emph{t})]+\emph{C}\emph{T}, and \emph{C} = ln(\emph{B}\emph{$\omega$}\emph{a}/2$\pi$\emph{r}) is assumed as a constant, where \emph{B} is the magnetic induction, $\omega$ is the attempt frequency for vortex hopping, \emph{a} is the hopping distance, and \emph{r} is the sample radius. By selecting \emph{C} = 18, all the curves can be scaled together as shown in Fig. 4(b). The solid line indicates power-law fitting by Eq. (3) to the large \emph{J} region where the slope in Fig. 4(a) is positive. Deviation of the data from the fitting line in the small \emph{J} region is reasonable since vortex creep is plastic there. The fitting gives glassy exponent $\mu$ = 1.27, activation energy \emph{U}$_{00}$ = 22 K, and \emph{J}$_{c00}$ = 2.5 $\times$ 10$^5$ A/cm$^2$. The value of glassy exponent obtained from the extended Maley's method is very close to that evacuated in Fig. 4(a). With the value of \emph{U}$_{00}$, temperature dependence of \emph{S} is fitted by Eq. (2) with a single free parameter of $\mu$ln(\emph{t}/\emph{t}$_{eff}$) = 35 shown as the dashed line in Fig. 3(b).

Using the parameters obtained above, we can calculate \emph{J} after flux creep from Eq. (4), which is shown as the solid line in Fig. 5. Obviously \emph{J} is reasonably reproduced except for the deviation at 2 K, which comes from the self-field effect as already discussed above. The good match between experimental results and the calculation means that the present analyses based on the collective creep theory is appropriate. To get more insight into the pinning mechanisms in FeTe$_{0.6}$Se$_{0.4}$, we also calculated \emph{J} without flux creep using the generalized inversion scheme (GIS) \cite{37,38}. Although in this scheme, we have to assume empirical temperature dependence of penetration depth $\lambda$ and coherence length $\xi$ as $\propto$(1-\emph{t}$^4$)$^{-1/2}$ and $\propto$(1+\emph{t}$^2$)$^{1/2}$(1-\emph{t}$^2$)$^{-1/2}$, respectively, we can directly reconstruct the true critical current density \emph{J}$_c$ without creep from \emph{J}$_s$ and can discuss the pinning mechanism. To compare with the theoretical prediction, we simply choose parameters for the three-dimensional single vortex pinning, and assume ln(\emph{t}/\emph{t}$_{eff}$) $\sim$ 28 in consistent with the analysis in Fig. 3(b). \emph{J}$_c$ is reconstructed as presented in Fig. 5. Shown together with \emph{J}$_c$ are the theoretical predictions for $\delta$\emph{l} pinning \emph{J}$_c$(\emph{t})/\emph{J}$_c$(0) = (1-\emph{t}$^2$)$^{5/2}$(1+\emph{t}$^2$)$^{-1/2}$ and for $\delta$\emph{T}$_c$ pinning \emph{J}$_c$(\emph{t})/\emph{J}$_c$(0) = (1-\emph{t}$^2$)$^{7/6}$(1+\emph{t}$^2$)$^{5/6}$ \cite{39}. Obviously, \emph{J}$_c$ resides between the predictions of $\delta$\emph{l} and $\delta$\emph{T}$_c$ pinning. Thus, the $\delta$\emph{l} pinning associated with charge-carrier mean free path fluctuations, and the $\delta$\emph{T}$_c$ pinning associated with spatial fluctuations of the transition temperature, coexist in FeTe$_{0.6}$Se$_{0.4}$ single crystal. Such a result is similar to those reported in Co-doped \cite{11,12} and K-doped \cite{40} BaFe$_2$As$_2$. However, it should be noted that dominance of $\delta$\emph{l} pinning in FeTe$_{0.7}$Se$_{0.3}$ \cite{18} and FeTe$_{0.5}$Se$_{0.5}$ are reported \cite{19}. These conclusions may come from the underestimated critical current density because of the sample quality (lower \emph{T}$_c$ \cite{18} and smaller \emph{J}$_s$ \cite{19}) and the insufficient lowest measuring temperature of $\sim$ 0.4 \emph{T}$_c$. More importantly, direct comparison of the theoretical curve with the experimental \emph{J}$_s$ after flux creep is also inappropriate.

In summary, we have studied the vortex dynamics in FeTe$_{0.6}$Se$_{0.4}$ single crystal by magnetization measurements and its time relaxation. Sharp superconducting transition width, large critical current density and homogeneous current distribution revealed by MO imaging manifest the very high quality of the crystal. Temperature dependence of \emph{S} shows a plateau in the intermediate temperature region with a high vortex creep rate \emph{S} $\sim$ 0.03, which is further interpreted by the collective creep theory. A crossover from elastic to plastic creep regime was also observed. With an extended Maley's method, the glassy exponent and pinning energy were directly determined as $\mu$ = 1.27 and \emph{U}$_{00}$ = 22 K at \emph{H} = 10 kOe. With these parameters, we successfully reproduce the screening current density obtained from magnetic hysteresis loops, which is actually after flux creep. The generalized inversion scheme was applied to analysis the critical current density without flux creep, which proves that the $\delta$\emph{l} and $\delta$\emph{T}$_c$ pinning coexist in FeTe$_{0.6}$Se$_{0.4}$ single crystals.

\acknowledgments
This work was partly supported by the Natural Science Foundation of China, the Ministry of Science and Technology of China (973 project: No. 2011CBA00105), and the Japan-China Bilateral Joint Research Project by the Japan Society for the Promotion of Science.

\vspace{1cm}
$^{*}$zxshi@seu.edu.cn   $^{\dag}$tamegai@ap.t.u-tokyo.ac.jp

\end{document}